**Origin of conductivity cross over in entangled multi-walled carbon nanotube network filled by iron**

George Chimowa[1,3], Ella C. Linganiso[1,2,3], Dmitry Churochkin[1], Neil J. Coville[2,3], and Somnath Bhattacharyya[1,3*]

[1]*Nano-Scale Transport Physics Laboratory, School of Physics,* [2]*Molecular Sciences Institute, School of Chemistry, and* [3]*DST/NRF Centre of Excellence in Strong Materials, University of the Witwatersrand, Private Bag 3, WITS 2050, Johannesburg, South Africa*

A realistic transport model showing the interplay of the hopping transport between the outer shells of iron filled entangled multi-walled carbon nanotubes (MWNT) and the diffusive transport through the inner part of the tubes, as a function of the filling percentage, is developed. This model is based on low-temperature electrical resistivity and magneto-resistance (MR) measurements. The conductivity at low temperatures showed a crossover from Efros-Shklovski (E-S) variable range hopping (VRH) to Mott VRH in 3 dimensions (3D) between the neighboring tubes as the iron weight percentage is increased from 11% to 19% in the MWNTs. The MR in the hopping regime is strongly dependent on temperature as well as magnetic field and shows both positive and negative signs, which are discussed in terms of wave function shrinkage and quantum interference effects, respectively. A further increase of the iron percentage from 19% to 31% gives a conductivity crossover from Mott VRH to 3D weak localization (WL). This change is ascribed to the formation of long iron nanowires at the core of the nanotubes, which yields a long dephasing length (e.g. 30 nm) at the lowest measured temperature. Although the overall transport in this network is described by a 3D WL model, the weak temperature dependence of inelastic scattering length expressed as $L_\phi \sim T^{-0.3}$ suggests the possibility for the presence of one-dimensional channels in the network due to the formation of long Fe nanowires inside the tubes, which might introduce an alignment in the random structure.

______________________________________________________________________________

*Somnath.Bhattacharyya@wits.ac.za

## I. INTRODUCTION

Since the discovery of carbon nanotubes (CNTs) much attention has been given to the study of the device characteristics of this material by many researchers due to their interesting electrical and magnetic properties [1-5]. Metal filled multi-walled carbon nanotubes (MWNTs) may find several important applications ranging from magnetic sensors made of a single (metal filled) tube, scanning probe microscopy tips to the assembly of aligned high density magnetic nano-cores for future magnetic data storage [2]. Tuning the transport properties in metal filled MWNTs as a function of the magnetic field will be very important in band gap design engineering and allow for the development of real molecular level single electron devices as well as devices for spintronics. As such, the combination of magnetic materials and CNTs having excellent electrical transport properties becomes an essential research direction. A single multi-walled nanotube shows ballistic transport over a long length. However, a wide variation in the transport properties of an entangled MWNT network, from the hopping to the diffusive regime has been reported, depending on the degree of disorder in the network [3, 4]. Experimental work on the Aharonov–Bohm resistance oscillations in MWNTs has suggested that the electric current path in these materials is primarily along the outer most shell, perhaps because of the way the contacts are made in CNTs [5]. Our studies on the electron transport properties of Fe filled MWNTs, suggest the participation of the inner most shell in addition to the outer most shell. Metal incorporation in nanotubes can introduce defects, particularly in the inner shell furthermore, the metal islands formed inside CNTs can form multiple hopping paths or diffusive channels through the metal nanowires depending on the percentage of the metal incorporated therein. Reports by other researchers on metal filled MWNTs have indicated one conduction mechanism, either hopping or 3D weak localization, in the carbon nanotube network but the analysis of experimental data revealing both these two mechanisms has not appeared to date [6,7]. Magneto-resistance (MR) studies by various authors on MWNT composites and empty metallic/semiconducting single walled nanotubes (SWNTs) [8-10] in the VRH regime have shown that quantum interference between many possible hopping paths in a magnetic field would lead to a pure negative MR, which may be parabolic at low field and linear at higher fields. In previous studies, which focused on low to medium magnetic fields, the connection between conductivity crossover and the filling of the MWNTs was not clearly portrayed [11]. A clearer picture on how the filling affects electrical conductivity can be expected from low temperature transport that combines the effect of high magnetic fields (i.e. 12 T) with the variation in the amount of metal filling. Herein, we show that the conductivity crossover in iron filled MWNTs, is dependent on the filling content. The weak inter shell coupling in MWNTs defines two distinct conduction channels, namely the outer most shell, which links tubes in the network and the inner shell, which links the metal nanowires at the core. The introduction of iron can ideally enhance the participation of the inner most shell in the conductivity where the dephasing time variation with temperature can be controlled. This confirms that the system remains in the weak scattering limit for the highly filled MWNTs. From the analysis of data we have established a temperature dependence on the dephasing length. These observations are explained with a model, emphasizing the contribution from both the inner shell and the outer shell on the conductivity of MWNTs. The model is consistent with microstructural studies of Fe-CNT samples [13] and can be used to develop a class of fast electronic devices.

## II. EXPERIMENT

Ferrocene was used as a precursor to supply both the metal for filling the nano-tubes and the carbon source for the growth of nano-tube shells. Dichlorobenzene was used as a supplementary source of carbon. About 1.0 g of ferrocene was loaded into a quartz boat and positioned in the first temperature zone of a two zone quartz tube reactor. The deposited material was collected in the second temperatures zone of the quartz tube which was used as a substrate. Ferrocene was passed through a bubbler containing dichlorobenzene and sublimated under a flow of a mixture of 5 % $H_2$/Ar (300 sccm) in the first temperature zone. The sublimation temperature was kept at 175 $^o$C while the deposition of filled carbon nanotubes on the quartz reactor walls was accomplished at 900 $^o$C for 30 min [12]. The reactor was cooled to room temperature under a flow of argon and the samples were treated with hydrochloric acid to remove iron on the tube's exterior surface and then characterized using various techniques. These included transmission electron microscopy (TEM), thermogravimetric analysis (TGA) and X-ray diffraction (XRD) [13]. Four samples containing different amounts of iron were chosen for detailed study and are represented as 3%Fe-MWNT, 11% Fe-MWNT, 19% Fe-MWNT and 31% Fe-MWNT where the percentage of iron relates to the iron content of the CNTs. The Fe-MWNTs were contacted on the four corners of a rectangle with a highly conducting silver paste (RS-186-3593) see Table I for the exact dimensions. The temperature dependence of the resistance (*R-T*) and MR were measured in the temperature range 2.5 K – 150 K and at fields up to 12 T on a completely automated cryostat (Cryogenic Ltd., UK).

### III. RESULTS AND DISCUSSION

TEM images of the iron filled MWNTs [Fig. 1(a) and (b)] revealed that the as grown Fe-MWNT nanocomposites consist of a mixture of nanowires, nanorods and nano-particles of iron inside the CNTs; the filling of iron was not continuous. The outer diameter of the CNTs varied between 11 nm and 45 nm. At low Fe loading the tubes are thoroughly entangled whereas at higher iron loading a tendency to align by the metal filled tubes can be noticed [Fig. 1(b)]. A detailed microstructure analysis showed that the inner diameter of the tubes (i.e. the diameter of the iron rods) can be 10 nm or less while the length of the rods can be extended over several hundreds of nanometers. An increase in Fe content in the tubes generates tubes that are straight. High-resolution TEM micrographs showed that the metal nanowires are tightly bound to the inner wall of the MWNTs [13].

## A. Temperature dependence of conductivity

The resistance-temperature (*R-T*) dependence measurements clearly show the electrical transport to be dependent on iron content (Fig. 2). From the graph a decrease in the semiconducting behavior with an increase in iron content is observed below 40 K. The sample with the least iron content (3%) shows a sharp increase of resistance with decreasing temperature such that it was highly insulating at lower temperatures and could not be measured with our system. The 31% sample shows a reduced semiconducting or semi-metallic behavior. We have investigated the origin of the conductivity increase as a function of iron percentage in the samples and suggest, firstly, the increased probability of electron transfer between entangled neighboring tubes in the network results in improved conductivity. Secondly the Fe provides an alternative conduction path inside the core of the nanotubes. Large iron clusters can result in weak localized (diffusive) transport in these systems. These basic concepts are verified from proper fitting of *R* vs. *T* data given below.

### 1. Strongly localized conduction

Analysis of the individual *R-T* measurements reveals that the conduction process is predominately hopping as suggested by the Mott and Efros-Shklovskii (E-S) models, Eq.(1).

$$R_T = R_0 \exp [(T_0/T)^{1/d+1}] \qquad (1),$$

where $T_0$ in the E-S model is defined as $T_0 = \beta e^2/k_B \kappa \lambda$. Here $\kappa$, $\lambda$, and $k_B$ correspond to the dielectric constant of the material, electron localization length, and Boltzmann constant, respectively. In 3D the exponent *d* is equal to 2 (for the E-S model) and the numerical coefficient $\beta$ can have a value of 2.8 [14]. In the Mott model *d* is equal to 3 and $T_o = 18/k_B \lambda^3 N(E_F)$, where $N(E_F)$ represents the density of states at the Fermi level [15].

It has been predicted theoretically that when the metal content in CNTs is low, Coulomb interactions between electrons of distant metallic CNTs induce a soft Coulomb gap in the density of states of conduction [16]. This results in a E-S VRH conduction mechanism. An increased metal content enhances the screening of the Coulomb potential [8], which results in a weakening of Coulomb interactions, and the conduction mechanism changes from E-S VRH to the conventional Mott VRH. A $T^{-1/2}$ dependence of ln$R$ can sometimes be taken to mean 1D VRH in the Mott model. However, for our samples this is unlikely since the CNTs consist of networks of unaligned tubes [see Fig. 1(a)]. Fig. 3(a) indicates that both 3% and 11% Fe-MWNT samples can be described by the E-S model while the 19% and 31%Fe-MWNT samples deviate from the model as the temperature increases [see Fig. 3(b)]. Figure 4 shows that the 3D Mott VRH is a better model for the 19%Fe-MWNT samples. The low temperature region of the 31% Fe-MWNT sample in Fig. 4 shows a linear trend in the Mott VRH fit, but at higher temperatures pronounced deviation occurs. This could be due to some additional non-hopping contribution for the 31% Fe-MWNT sample. From the VRH fittings we are able to estimate the E-S and Mott characteristic temperatures denoted by $T_{\text{Mott}}$ and $T_{\text{E-S}}$, respectively (see Table II). MR measurements in the section III.B provide additional parameters, such as the localization lengths ($\lambda$) from which we calculated the hopping ranges ($R_h$) using equations (2) and (3). The data is given in table II.

$$R_{h.Mott} = \frac{3}{8}\left(\frac{T_{Mott}}{T}\right)^{1/4}\lambda \qquad (2),$$

$$\text{and } R_{h.E-S} = \frac{1}{4}\left(\frac{T_{E-S}}{T}\right)^{1/2}\lambda \qquad (3).$$

*2. Weakly localized conduction*

As indicated earlier, when the iron content is increased to 31% in the MWNT samples a clear deviation from the hopping conduction is noted [Fig. 4], which can be corrected by addition of a weak localization (WL) term to the conductivity. We attempted to obtain a fit to *R-T* data based on 2DWL and on 1DWL models for the 31%Fe-MWNT samples. However, a strong discrepancy from the data was observed. We have thus used a general approach to determine the conductivity correction due to the localization effects combined with the effect of *e-e* interactions in 3D. In this approach the conductance (*G*) is defined as a sum of Drude conductance ($G_0$) and the correction terms:

$$G_{3D}(T) = G_0 + \frac{e^2}{h\pi^3}\frac{1}{a}T^{p/2} + m\,T^{1/2} \qquad (4),$$

where *a* and *m* correspond to proportionality constants for WL and electron-electron scattering terms, respectively. The exponent *p* depends on the scattering mechanism and dimensionality of the system [18].

The conductance *vs.* temperature (*G-T*) graph in Fig. 5(a) shows the validity of a 3D WL model to describe the conduction in 31% Fe-MWNT samples. However this model is not appropriate for the 3% and 11% Fe-MWNT samples [see Fig. 5(b)]. Using Eq. (4) the exponent *p* is found to be about 0.6 for the 31% sample. The importance of the *p* value in the weak localization theory comes from the fact that it identifies unambiguously the dimensionality and the dominating mechanism of dephasing. Indeed, in 3D systems the theoretical values of the exponent '*p*' appeared to be equal to 3/2, 2 or 3 depending on the type of scattering employed, these values correspond to *e-e* scattering in the dirty limit, clear limit and a domination of electron-phonon scattering over inelastic scattering rate, respectively. This approach did not yield a proper fit of the data with these values of *p* (theoretical). In 2D, we may expect a $T^1$ trend which is determined by *e-e* scattering. On the contrary, the detailed *G* vs. *T* analysis, along with the MR analysis (see next section) gives a value of *p* significantly less than unity. Moreover, the exact value of *p* was around 0.6 for the 31% sample which, as we assume, has a most pronounced WL trend. Notice that the theoretically predicted values for *p* in a 1D system, due to *e-e* interactions is about 2/3 i.e. 0.66 [18]. The close proximity of the observed values for *p* with the mentioned theoretical value allows us to make a conjecture about 1 D structural features, which emerge in a Fe-filled entangled carbon nanotube network.

TEM analysis of the Fe-CNT samples show rod-like structures due to the iron inside the tubes, which penetrate throughout the sample [Fig. 1(b)]. In our opinion, this means that the structure of the elementary conduction units can be considered as 1D, leading to 1D effective dephasing by *e-e* interaction inside the unit. In comparison to the dephasing processes, overall conductivity as an averaged transport characteristic of the whole 3D sample would be structurally insensitive, and may still demonstrate the 3D WL and 3D *e-e* interaction contributions in accordance with Eq.(4).

## B. Magnetic field dependent conductivity

### *1. MR in strongly localized systems*

The MR measurements recorded for the iron filled MWNTs show a conductivity crossover from negative to positive MR between 2.5 K to 10 K as magnetic field increases [Figs.6 (a), (b) & (c)]. This crossover tendency is suppressed as the iron content and temperature is increased [Fig. 6 (d)]. At low temperatures the electron orbitals shrink, thus reducing the hopping probability, resulting in a positive MR. The negative MR is a result of quantum interference in the hopping regime or an effect due to WL in the metallic nanowires or metallic shells [19]. At low temperatures the wave function shrinkage and quantum interference mechanisms are additive with the former being more pronounced at high fields. As the temperature increases the wave-function shrinkage becomes suppressed and thus the MR becomes negative. We however, still observe a small contribution from orbital shrinkage for the 11% Fe-MWNT sample as evidenced by the small upturn in the MR at 50 K [Fig. 6(d)].

The two additive mechanisms affect the conductivity differently and this was modeled as

$$\ln[(\rho_H/\rho_o)] = -a_1 B + a_2 B^2 + a_3 \qquad (5),$$

where the coefficients $a_1$, $a_2$, and $a_3$ account for the quantum interference mechanism, the wavefunction shrinkage, and for the complex nature of hopping at low fields, respectively [3, 20]. Fitting the MR results with Eq. (5), [solid lines in the insert of Fig. 6 (b)], enables us to evaluate the coefficients $a_1$ and $a_2$, whose temperature dependence consequently gives additional information about the hopping and scattering processes in the samples. In strongly localized materials the positive MR is described by wavefunction shrinkage as

$$\ln[(\rho_H/\rho_0)] = t(\lambda/L_H)^4 (T_{mott}/T)^{3/4} \qquad (6),$$

where $t$ is a constant equal to 0.0025 in 3D and $L_H = (\hbar/eB)^{1/2}$ represents the magnetic length [11]. For the E-S model the exponent in Eq.(6) is changed to 3/2. Using Eq.(6) and the temperature dependence of $a_2$ [see Fig. 7(a) and (b)] we calculated the electron localization lengths, shown in Table II.

The coefficient of the linear term ($a_1$) obtained from the fitting of the MR results can be plotted as a function of temperature [Fig. 7(c)]. In the strong scattering limit, $a_1$ is theoretically expected to be constant and it can increase as temperature decreases in the weak scattering limit [22]. The latter is in agreement with our results presented in Fig.7(c) where $a_1 \sim T^{-0.68}$. A theoretical $T^{-7/8}$ dependence in 3D systems is predicted [3]. The exponent of $T$ is sensitive to the nature of disorder and dimensionality. The present analysis shows that the transport in these entangled Fe-MWNT networks containing low percentages of Fe, is in the weak scattering limit with a long characteristic length of about 5 nm.

## 2. MR in weakly localized systems

We now turn our attention to the magneto-conductance (MC) data to verify the 3D WL observed in the 31% Fe-MWNT sample. The theory of MC due to 3D WL formulated by Kawabata [23] is governed by

$$\Delta \sigma_{WL}(B,T) = \frac{e^2}{2\pi^2} \frac{1}{\hbar l_B} F(\delta) \qquad (7),$$

where the magnetic length $l_B = \sqrt{\hbar/eB}$. The function F($\delta$) [where $\delta = l_B^2/4L_\phi^2$, $L_\phi$ being the inelastic scattering length] has two limits for an analytical solution of the form $\Delta\sigma_{WL}$ proportional to $B^n$ ($B$ in Tesla) given by

$$\Delta\sigma_{WL}(B,T) = C\sqrt{B} \text{ for } \delta \ll 1 \qquad (8),$$

$$\text{and } \Delta\sigma_{WL}(B,T) = CB^2 \text{ for } \delta \gg 1 \qquad (9).$$

The condition in Eq. (8) {where $\delta \ll 1$} is observed for high fields at low temperatures. The square root of $B$ behavior of the conductivity is independent of system parameters, is a behavior unique to 3D WL [23]. Figure 8(a), inset clearly shows a $B^{½}$ dependence of the magneto-conductance at low temperatures and high magnetic fields. This $B^{½}$ dependence is not applicable at high temperatures and low fields instead a $B^2$ dependence at low fields can be seen. We have verified the invalidity for a 2DWL fit based on the $G$-$T$ data and have checked the suitability for a 1D WL model in this system [Fig. 8(b)]. The MR data, based on 1DWL, was fitted with Eq. (10), expressed as

$$\Delta G_{1DWL}(B,T) = -\frac{2Ne^2}{Lh}\left[\frac{1}{L_\phi^2} + \frac{1}{D\tau_B}\right]^{-0.5} \qquad (10),$$

where $\tau_B = \frac{3\ l_B}{W^2 D}$. According to the 1DWL model the MR depends on the radius ($W$) of the channel (i.e., average inner radius of the nanotubes), the length ($L$) of the channels (tubes) and the number of channels ($N$) participating in the conduction. The evaluation of the experimental data on the basis of a 1D MR approach shows that the proper fit requires at least $10^3$ conducting channels, i.e. iron-filled nanotubes, with an average width of 10 nm and the effective length of about 100 nm at a temperature of 5 K. For the 31% Fe filled sample with a cross-section of $2\times10^{-7}$ m$^2$ an estimated maximum number of conduction channels $\sim10^9$ were evaluated. The evaluation was for ideally aligned closely packed nanotubes. In practice, we may expect the number of channels to be $10^6$ times less than the ideal case because of the entanglement of nanotubes in the sample. This finds a good agreement with experiment. Changing the temperature should influence both the effective number of channels and their length as was observed in our measurements. A reasonably good fit to data using the 1D WL model is observed, which enabled us to determine the temperature dependence of the dephasing time ($\tau_\phi$). This dependence is found to be close to the theoretically defined one given by $\tau_\phi = \left(\frac{e2\sqrt{D}k_B R}{\hbar^2 L}\right)^{-2/3}$, where $R$ represents the resistance of a 1D sample per unit length [see Fig. 9]. A marked similarity is found between the experimentally derived $L_\phi(T) \sim T^{-0.25}$ value and the theoretically predicted one (i.e. $T^{-0.33}$).

Finally, the effect of 3D WL to conductivity through the fitting of MR data and calculation of the temperature dependence of the dephasing length ($L_\phi$) was achieved [Fig. 8(a) and 9]. The value of $L_\phi$ is relatively long (about 30 nm at 2.5 K), which is comparable to the previously reported value for a single and unfilled MWNT [17, 21]. It is important to notice that the exponent for $L_\phi(T)$ is 0.3 a value that is very similar to that of empty MWNTs. These features establish the effect of the disorder induced *e-e* interaction in 1D systems [21]. To calculate the dephasing time ($\tau_\phi$) in this system the value of the diffusion coefficient (*D*) was taken as 1 cm$^2$/s, which gives $\tau_\phi$ in a range of 10 ps. A much larger value of *D* has been reported earlier for this system [24]. At a high level of filling the transport properties are defined mostly by the internal content of the nanotube, i.e. the confined iron, which has a diameter of ~10 nm and length over several hundreds of nanometers. In our opinion, the confinement is mainly reflected in the temperature dependence of $\tau_\phi$ which is, as is well-known, very sensitive to the dimensionality of the corresponding dephasing mechanism [13]. Thus, the presence of the iron in the nanotube introduces an additional characteristic scale, i.e. the inner diameter of a nanotube, which is about 10 nm, into the system. We assume that dephasing by *e-e* interactions mainly occurs inside the channels (nanotubes). Both the dephasing length and the length of the iron wires are larger than the CNT diameter, which reveals an effective one-dimensional network. This process gives the corresponding value of the temperature exponent for a dephasing time of 0.6, which is very close to the typical 1D value of 0.66 [see Fig. 9, inset]. Overall, the weak temperature dependence of $\tau_\phi$ can provide potential applications of this material as a fast electronic device [25].

To track the trend in the conductivity and MR in the present work we investigated Fe-MWNTs at the four different Fe fillings (3% to 31%). The conductivity for the 3% to 19% Fe-MWNT samples demonstrated a sharp increase with decreasing temperature, which we attributed to the hopping mechanism of conductivity. The 31% Fe specimen shows only smooth behavior in respect of temperature variation, which we ascribe to a 3D weak localization accompanied by the *e-e* interaction. The detailed analysis for the 3% and 11%Fe-MWNT samples revealed a hopping exponent of ½ whereas for the 19% sample the value was ¼. These observations allow us to suggest a conduction mechanism where the inner shell of the tubes, and iron at the core of the tubes, play a very important role as conduction pathways [see Fig. 10(a) to (e)]. The HRTEM study confirms the close coupling of the metal islands with the inner shell of the tubes [13]. The weak inter-shell coupling prohibits the hopping of conduction electrons across carbon nanotubes. The hopping of conduction electrons is less feasible from one iron site to another within the same tube for the 3% and 11%Fe samples since the small iron islands are located far from each other inside the tubes [see Fig. 10(a) and (b)]. However, the outer most shell provides the link between the carbon nanotubes within the network [Fig. 10(c)]. As mentioned above, the ½ in the exponent can be identified as a hallmark of the E-S hopping regime with a strong Coulomb interaction. However, increasing the iron concentration leads to the suppression of the Coulomb interaction and consequently, a crossover to the Mott regime with a ¼ exponent for the 19% samples. The localized states are situated around the outer sheath of the multi walled carbon nanotubes, which are tightly connected to the iron islands distributed mutually inside the connected ensemble of the carbon nanotubes [Fig. 10(d) and (e)]. The changing of the exponent in the hopping regime transport from ½ to ¼ with an increase of the iron filling from 11% to 19% can be explained through the increase of the screening of the localized states induced by the growth of the nearby iron islands. The rise in conductivity with increasing iron concentration up to 19% was interpreted as a spreading (elongation) of the iron islands, which promotes the higher probability of hopping. The hopping mechanism can explain the conductivity of the 11% and the 19% samples, which was supported by MR measurements. This mechanism also demonstrated pronounced negative character with a marked linear trend at high values of $B$ and $B^2$ behavior at low values. Such behavior is rather common for a hoping regime if one takes into account the mechanism proposed earlier i.e. interference between hopping paths, which leads to the negative MR, with exactly the same trends observed in both samples [26]. The positive tendency in the resistance at low temperatures was interpreted as a wave function shrinkage which is feasible at the values of magnetic field used. Thus, for the 11% and 19% samples it was established that the hopping approach worked well with the appropriate level of consistency for both conductivity and MR measurements. This model explains the reason for observing a nearly similar hopping distances in the 11%Fe and 19%Fe samples although the iron content is very different in these samples [see Table II].

In contrast, for the 31%Fe specimen, as was mentioned above, we are obliged to introduce a different proposal, which is based on interacting extended states influenced by a weak disorder. This inevitably leads to the 3D WL and electronic-electronic interaction corrections. Intuitively, this can be visualized as an appearance of almost continuous bulk iron domains [Fig. 10(d)]. Indeed, we observed a $B^{0.5}$ negative trend in MR at high magnetic fields, which is a trait of 3D weak localization. Moreover, the conductivity with a weak temperature dependence was also interpreted by Eq. (4) as 3D WL together with the 3D $e$-$e$ interaction. The nanotube specificity of the sample is probably reflected in the temperature dependence of $\tau_\phi \sim T^{-0.6}$ and is quite atypical for 3D WL phenomena. As the iron inside the tubes becomes continuous and extends over several nanometers the presence of 1D channels is revealed in the network [Fig. 10(e)].

## IV. CONCLUSIONS

In summary, conduction in filled MWNT networks with a low amount of Fe is dominated by hopping at very low temperatures. Introduction of metallic components into the tubes can result in some diffusive conduction due to weak localization effects. We have also shown that the introduction of iron results in a transition from Efros-Shkolovski to Mott variable range hopping. We have further suggested a mechanism based on the microstructure where two main conduction paths have been proposed using the inner most shell and the metal nanowires at the core of the nanotubes. We proposed that the weak inter shell coupling minimizes or prohibits the conduction across the multi-walls of the tubes. This, however, still needs further investigation to confirm this proposal. Magneto-resistance measurements revealed that at very low temperatures, in less filled carbon nanotubes, wavefunction shrinkage is the dominant process at high magnetic field. However, as the iron content is increased this process is suppressed and quantum interference of the electron's hopping paths becomes dominant. One of the major claims in this report is the weak temperature dependence of the dephasing length (whose value is also very long), which reveals 1D filamentary conduction channels, described theoretically in low-dimensional superstructures. We have also seen that as more iron is added to the nanotubes, weak localization processes begin to be noticed. This we believe is due to the increased density of metal nanowires at the core of the carbon nanotubes where a strong coupling between the metal and carbon is expected. This study will be useful in developing a new generation of devices based on metal filled carbon-nanotubes.

## ACKNOWLEDGEMENTS

S.B would like to thank the National Research Foundation (SA) for granting the Nanotechnology Flagship Programme support to perform this work and also the University of the Witwatersrand Research Council for financial support. We acknowledge the Electron Microscopy Unit for the use of the microscopes.

**Table I:** Sample dimensions used for electrical transport measurements.

| Sample | Length (m) | Width (m) | Thickness (m) |
|---|---|---|---|
| 3% Fe-MWNT | $5 \times 10^{-3}$ | $3 \times 10^{-3}$ | $5 \times 10^{-6}$ |
| 11% Fe-MWNT | $4.5 \times 10^{-3}$ | $3 \times 10^{-3}$ | $5 \times 10^{-6}$ |
| 19% Fe-MWNT | $5 \times 10^{-3}$ | $2.6 \times 10^{-3}$ | $5 \times 10^{-6}$ |
| 31% Fe-MWNT | $4 \times 10^{-3}$ | $3 \times 10^{-3}$ | $5 \times 10^{-6}$ |

**TABLE II**: Calculated parameters from both the *R vs. T* and MR measurements.

| Sample Fe % | $T_{E-S}$ | $T_{Mott}$ | Localization length $L_{E-S}$ | $L_{Mott}$ | Hopping Range $R_{E-S}$ | $R_{Mott}$ @ 5 K |
|---|---|---|---|---|---|---|
| 11% | 6.4 K | | 18.2 nm | | 5.15 nm | |
| 19% | | 4.85 K | | 14.8 nm | | 5.01 nm |
| 31% | | 1.22 K | | 17.9 nm | | 4.71 nm |

**References**

1. M. S. Dresselhaus, Nature (London) **358**,195(1992).

2. F. Geng and H. Cong, Physica B Cond. Matt. **382**, 300 (2006).

3. M. Jaiswal, W. Wang, K. A. S. Fernando, Y. P. Sun, and R. Menon, Phys. Rev. B **76**, 113401 (2007).

4. K. Liu, P. Avouris, and R. Martel, Phys. Rev. B **63**, 161404 (2001).

5. A. Bachtold, M. S. Fuhrer, S. Plyasunov, M. Forero, E. H. Anderson, A. Zettl, and Paul L. McEuen, Phys. Rev. Lett. **84**, 6082 (2000); C. Schönenberger, A. Bachtold, C. Strunk, J. P. Salvetat, and L. Forro, Appl. Phys A **69**, 283 (1999); Y. K. Kang, J. Choi, C. Y. Moon, and K. J. Chang, Phys. Rev. B **71**, 115441 (2005).

6. M. Aggarwal, S. Khan, M. Husain, T. C. Ming, M. Y. Tsai, T. P. Perng, and Z. H. Khan, Eur. Phys. J. B **60**, 319 (2007).

7. M. Baxendale, V. Z. Mordkovich, S. Yoshimura, R. P. H. Chang, and A. G. M. Jansen, Phys. Rev. B **57**, 24 (1998).

8. K. Yanagi, H. Udoguchi, S. Sagitani, Y. Oshima, T. Takenobu, H. Kataura, T. Ishida, K. Matsuda, and Y. Maniwa, ACS-Nano, **4**, 4027 (2010).

9. H. J. Li, W. G. Lu, J. J. Li, X. D. Bai, and C. Z. Gu, Phys. Rev. Lett. **95**, 086601 (2005).

10. Y. Z. Long, Z. H. Yin, and Z. J. Chen, J. Phys. Chem. C **112**, 11507 (2008); G. T. Kim, E. S. Choi, D. C. Kim, D. S. Suh, Y. W. Park, K. Liu, G. Duesburg and S. Roth, Phys. Rev. B **58**, 16064 (1998).

11. T. Takano, T. Takenobu and Y. Iwasa, J. Phys. Soc. Jpn. **77**, 124709 (2008).

12. R. Sen, A. Govindaraj, and C. N. Rao, Chem. Phys. Lett. **267**, 276 (1997).

13. E. C. Linganiso, G. Chimowa, P. Franklin, S. Bhattacharyya, and N. J. Coville (unpublished). E. C. Linganiso, MSc. Thesis (2010).

14. B. I. Shklovskii, and A. Efros, "Electronic properties of Doped Semiconductors"; Springer-Verlag:Berlin; pp 228-244 (1984).

15. N. F. Mott, "Conduction in non-crystalline Materials"; Oxford University Press: Oxford; pp 27-29 (1987).

16. T. Hu and B. I. Shklovskii, Phys. Rev B **74**, 054205 (2006).

17. N. Kang, J. S. Hu, W. J. Kong, L. Lu, D. L. Zhang, Z. W. Pan, and S. S. Xie, Phys. Rev. B **66**, 241403 (2002).

18. P. A. Lee and T. V. Ramakrishnan, Rev. Mod. Phys.**57**, 2 (1985).

19. O. Faran and Z. Ovadyahu, Phys. Rev. B **38**, 5457 (1988).

20. M. E. Raikh and G. F. Wessels, Phys. Rev. B **47**, 15609 (1993).

21. A. Bachtold, C. Strunk, J.-P. Salvetat, J.-M. Bonard, L. Forro, T. Nussbaumer, and C. Schönenberger, Nature (London) **397**, 673 (1999); R. Tarkiainen, M. Ahlskog, A. Zyuzin and M. Paalanen, Phys. Rev. B **69**, 033402 (2004).

22. N. V. Agrinskaya and V. I. Kozub, Phys. Stat. Sol. B **205**, 11 (1998).

23. A. Kawabata, J. Phys. Soc. Jpn. **49**, 2, 628 (1980).

24. J.-F. Dayen, T. L. Wade, M. Konczykowski, and J.-E. Wegrowe, Phys. Rev. B **72**, 073402 (2005).

25. S. Roche, F. Triozon, A. Rubio, and D. Mayou, Phys. Lett. **A 285**, 94 (2001).

26. V. L. Nguyen, B. Z. Spivak, and B. I. Shklovskii, Pis'maZh. Eksp. Teor. Fiz. **41**, 33 (1985) [JETP Lett. **41**, 42 (1985)]; Zh. Eksp. Teor. Fiz. **89**, 1770 (1985) [Sov. Phys.–JETP **62**, 1021 (1985)]; W. Schirmacher, Phys. Rev. B **41**, 2461 (1990).

**Figure captions:**

**Fig. 1** TEM images for the **(a)** 19% and **(b)** 31% Fe-MWNT samples.

**Fig. 2** Normalized *R-T* measurements for the 3%, 11%, 19% and 31% Fe-MWNTs.

**Fig. 3(a)** A ln*R* *vs.* $T^{-1/2}$ graph to evaluate the E-S model for 3% and 11%Fe-MWNT samples. **(b)** ln*R* *vs.* $T^{-1/2}$ fitting of 19% and 31%Fe-MWNT data shows a deviation from linearity at high temperature. The dotted lines are to guide the eyes.

**Fig. 4** A ln*R* *vs.* $T^{-1/4}$ graph shows the validity of Mott VRH model for the 19% Fe-MWNT sample and a deviation from the Mott 3D VRH model for 31%Fe-MWNT samples. The dotted lines are a guide to the eyes.

**Fig. 5(a)** A Conductance vs. temperature plot fitted with the 3D WL (red solid line for 31% and magenta solid line for 19% respectively) to check for WL contributions. Black dashed line, blue solid line and cyan solid line were plotted in order to show the pronounced deviation from the experimental data purely 3D *e-e* interaction, 2D WL and 1D WL correspondingly. **(b)** Separate plots for 3% and 11% Fe-MWNT samples show the deviation from the 3D WL fit. The faint dotted lines are a guide to the eyes.

**Fig. 6** Normalized MR measurements made at **(a)** 2.5 K, **(b)** 5 K, **(c)** 10 K, and **(d)** 50 K. The insert in **(b)** shows the natural log of the MR data for the 11%, 19% and 31% samples fitted with Eq.(5) as explained in the text.

**Fig. 7(a)** A graph of $a_2$ *vs.* $T^{-3/4}$ is plotted to determine the electron localization length for the 19% and 31% Fe-MWNTs. **(b)** A graph of $a_2$ *vs.* $T^{-3/2}$ is plotted to determine the localization length for 11% Fe-MWNTs. **(c)** A graph of $a_1$ *vs.* *T* is plotted to determine the scattering limit in the samples.

**Fig. 8(a)** A graph of magneto-resistance *vs*. B for the 31% Fe-MWNTs to show the applicability of the 3D WL model. **Insert:** MC *vs*. $B^{1/2}$ plot shows a linear fit. **(b)** Magneto-resistance *vs*. B for the 31% Fe-MWNTs fitted based on 1D WL model, the fits are the dotted lines.

**Fig. 9** Temperature dependence of the dephasing length obtained from both (3D WL and 1D WL) models. **Insert:** Temperature dependence of $\tau_\phi$ obtained from 3DWL fit to MR data.

**Fig. 10** The conduction mechanism in the Fe-MWNTs network is shown schematically. Case **(a)** the hopping range is far larger than the iron nanowire length. The conduction path is along the outer shell because of weak inter-shell coupling and hence 3D hopping to neighboring CNTs. **(b)** An increase of iron in the tubes results in the hopping range being comparable to the iron nanowire length and this allows for participation of the inner tube in the conduction process. **(c)** The entangled nanotube network is shown where hopping takes place between tubes. **(d)** As the iron content increases further transport is dominated by the weak localization effects due to inner shell conductivity and through the long iron nanowires in the partially aligned nanotubes. **(e)** Introduction of metal rods in the MWNTs can form an aligned structure and the probability for 1D conduction increases in the 3D network.

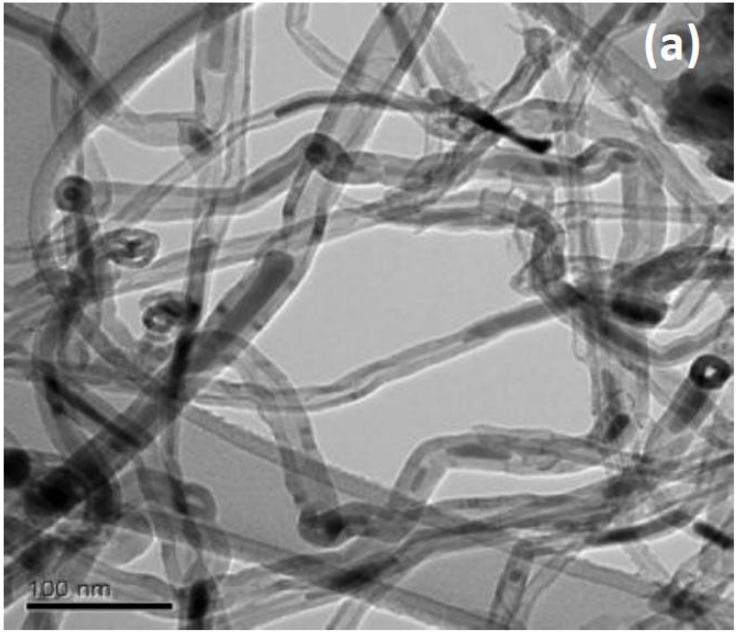

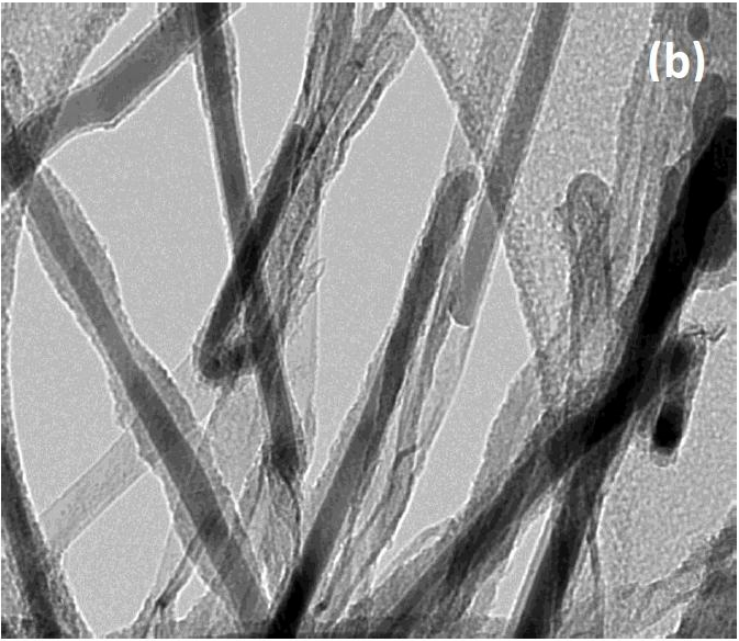

Fig. 1 (a) and (b) / Chimowa et al.

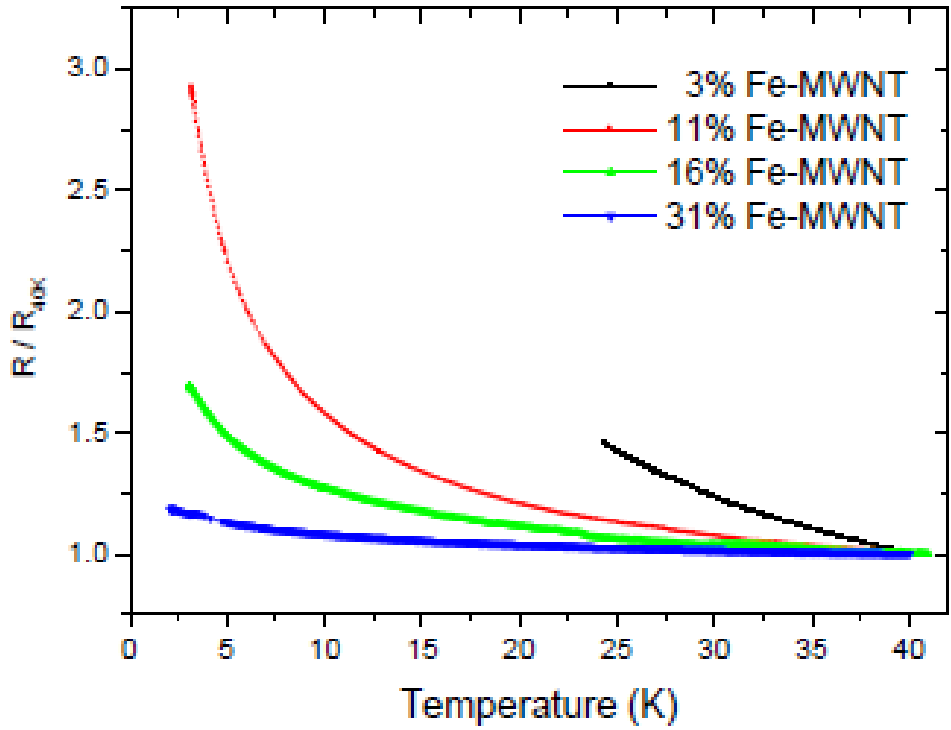

Fig. 2 / Chimowa et al.

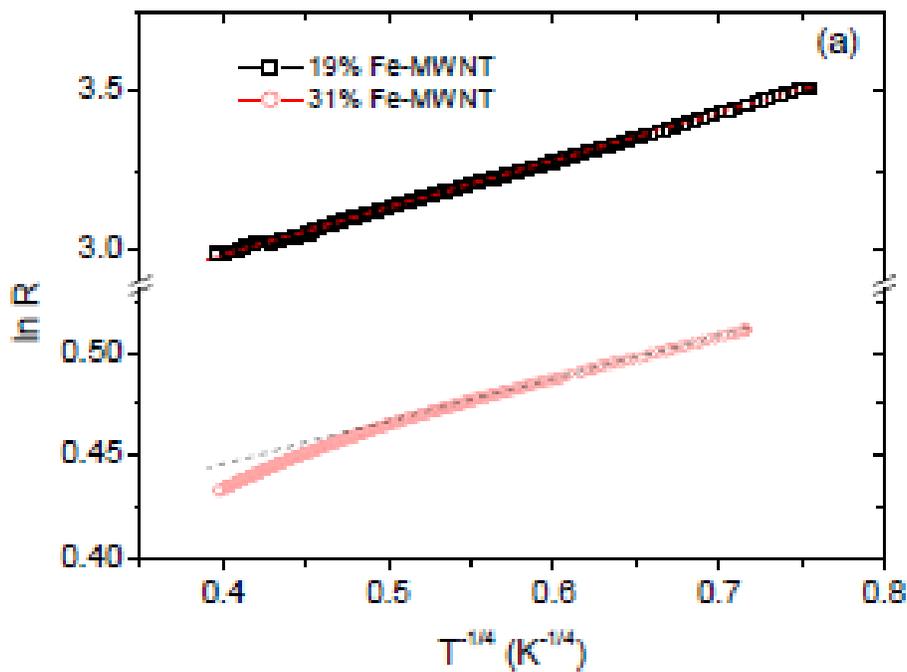

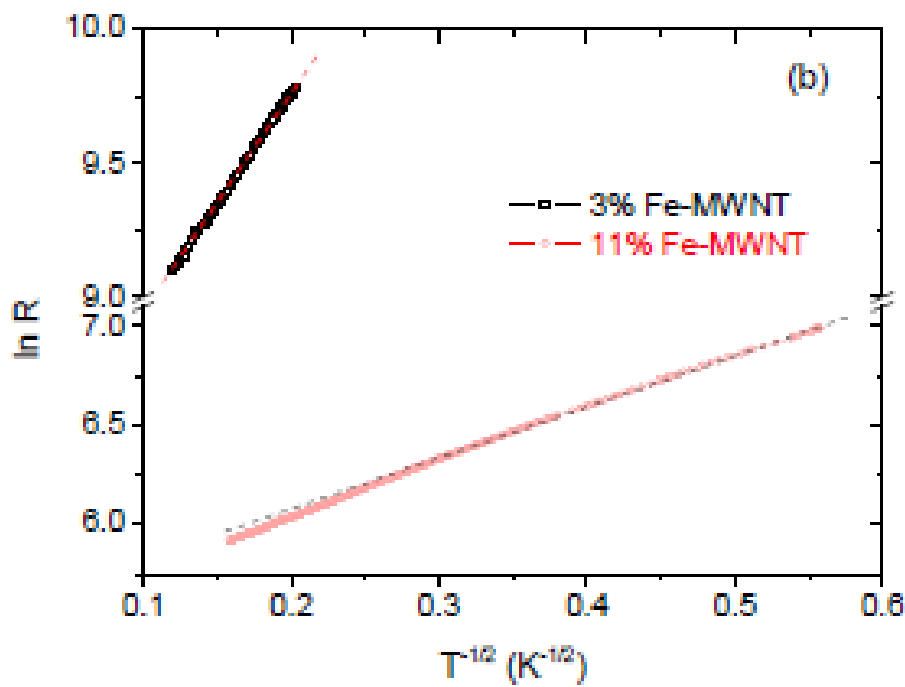

Fig. 3 (a) and (b) / Chimowa et al.

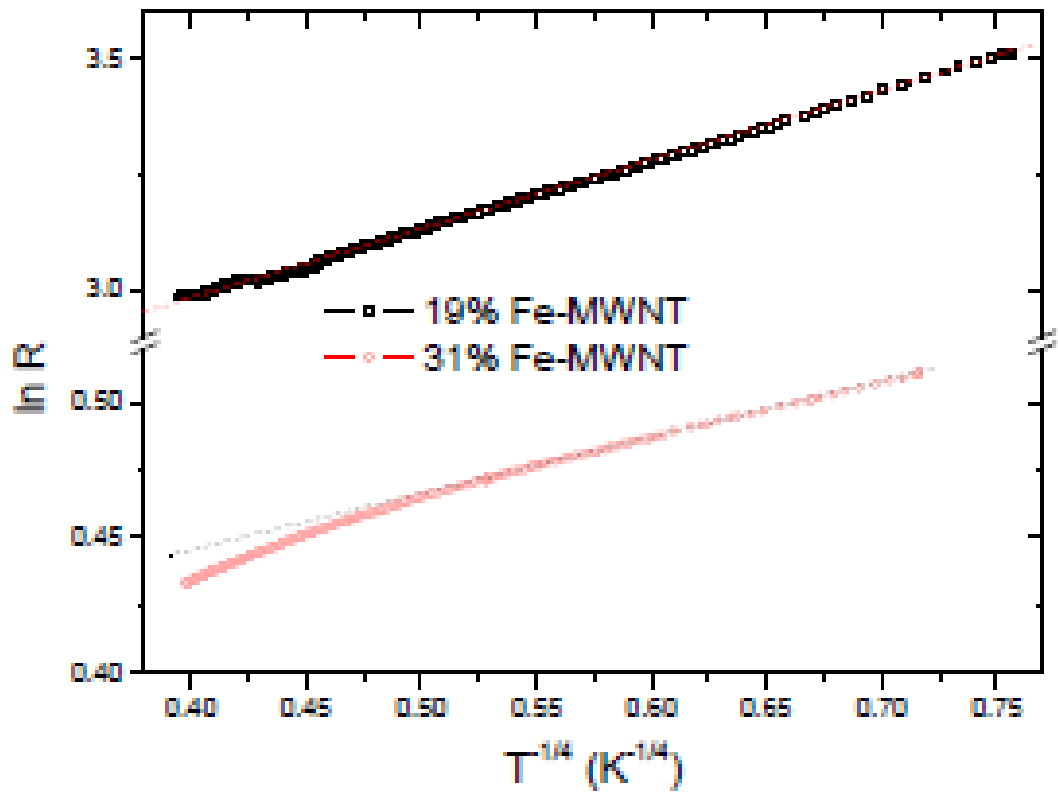

Fig. 4 / Chimowa et al.

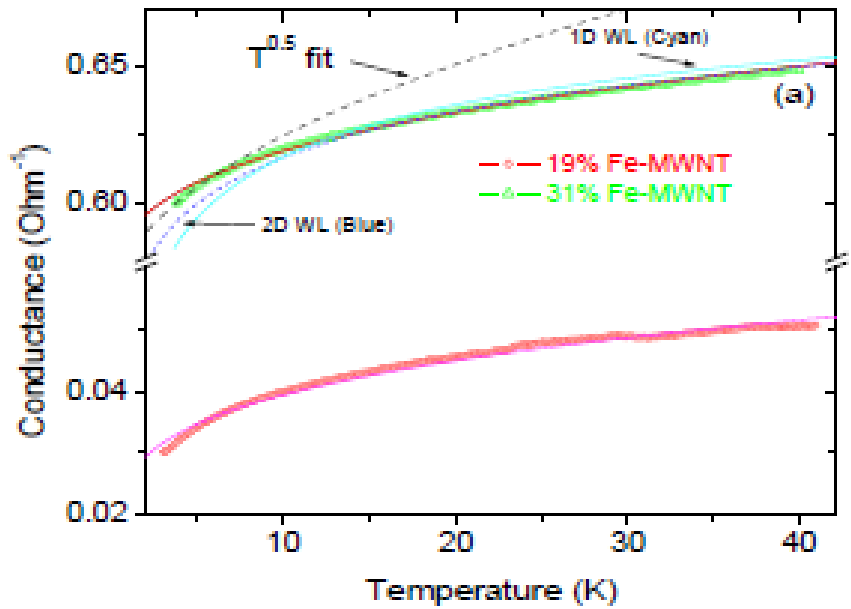

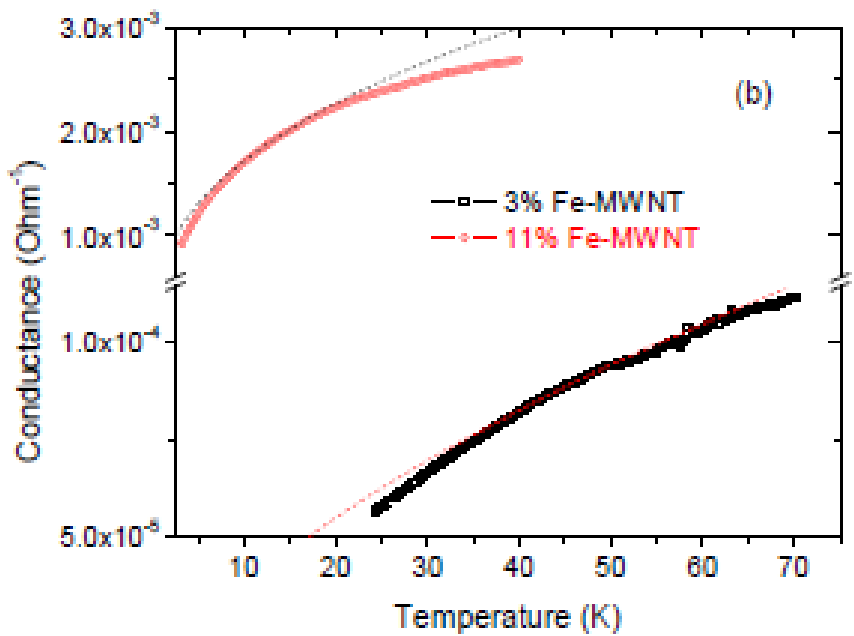

Fig. 5 (a) and (b) / Chimowa et al.

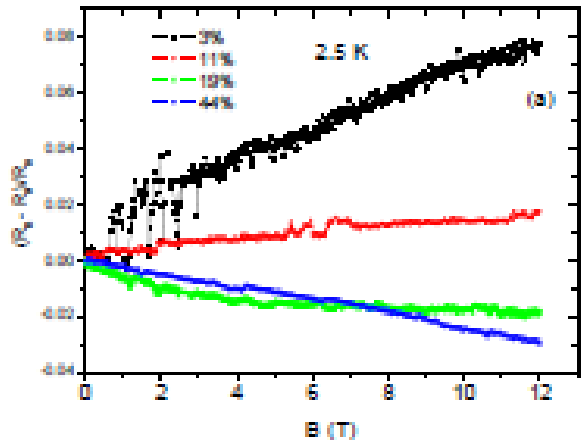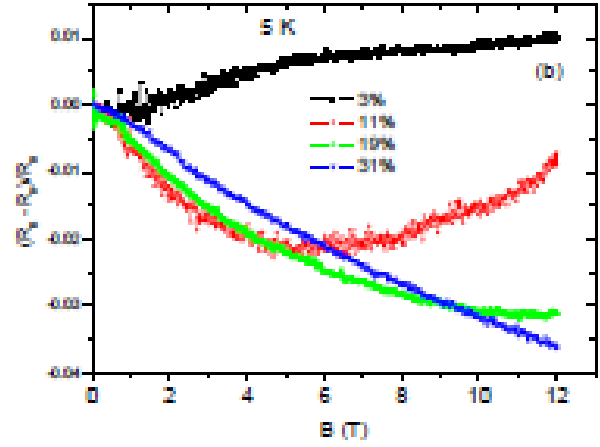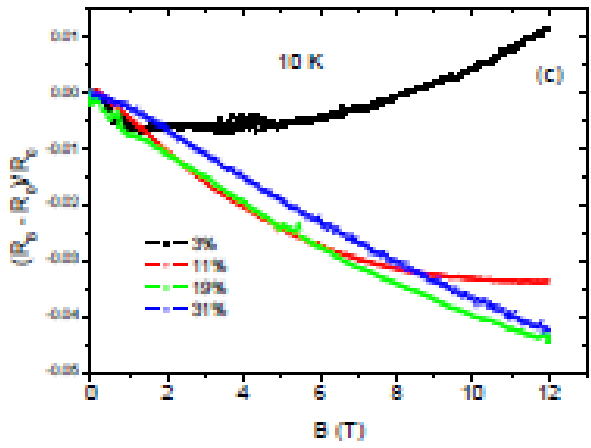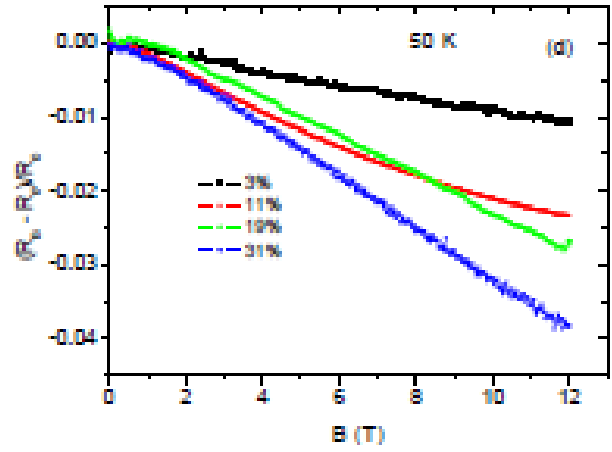

Fig. 6 (a) to (d) / Chimowa et al.

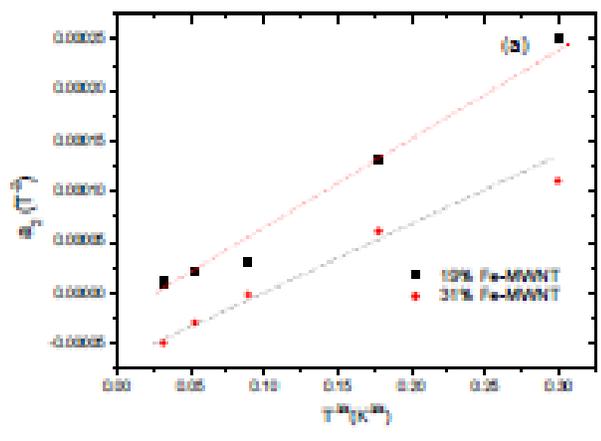

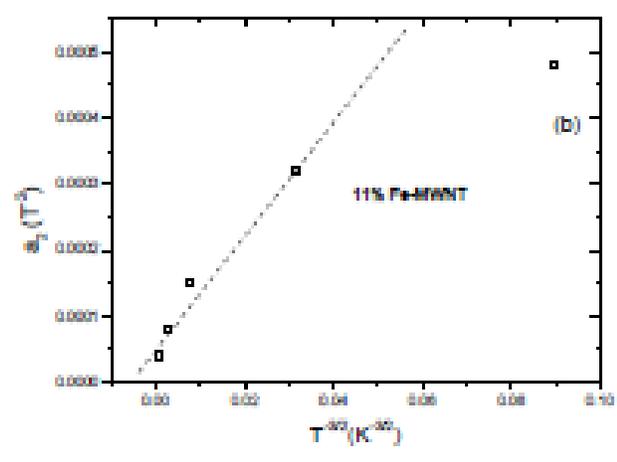

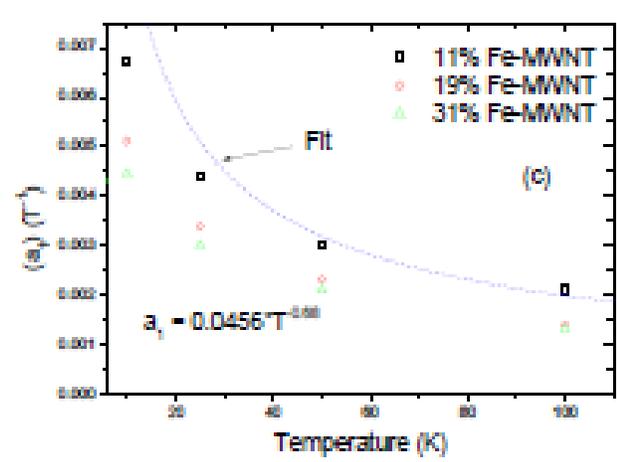

Fig. 7 (a), (b) and (c) / Chimowa et al.

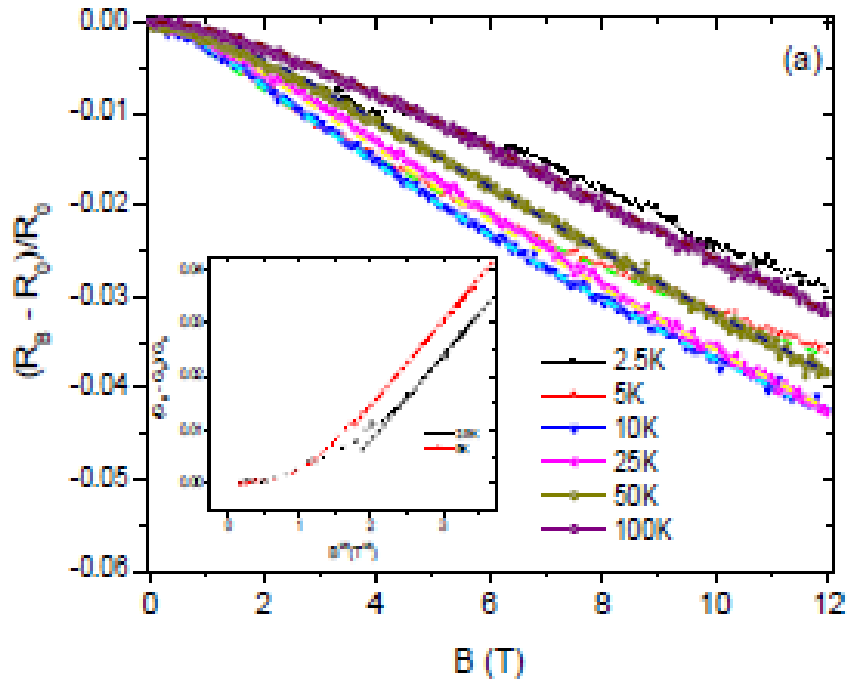

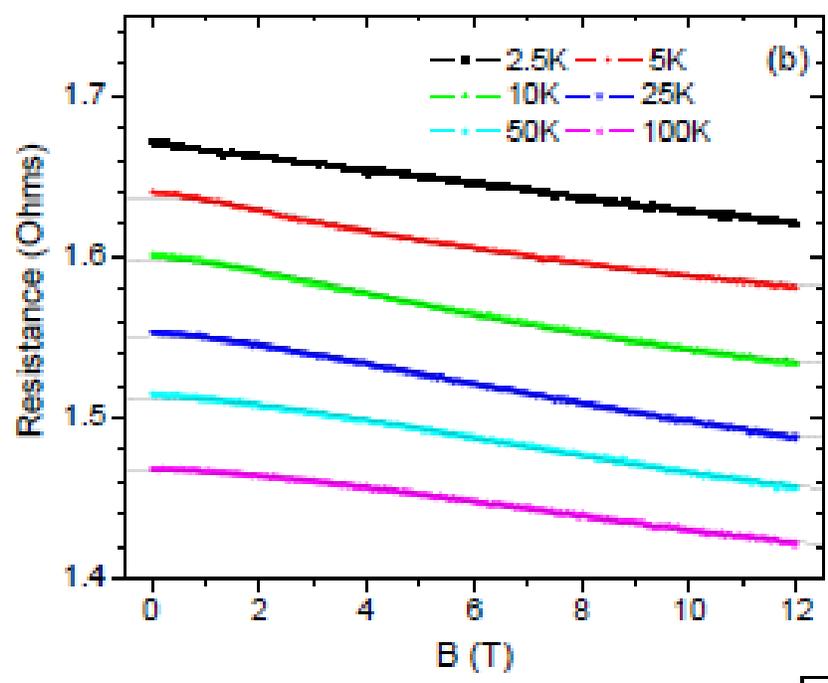

Fig. 8 (a) and (b) / Chimowa et al.

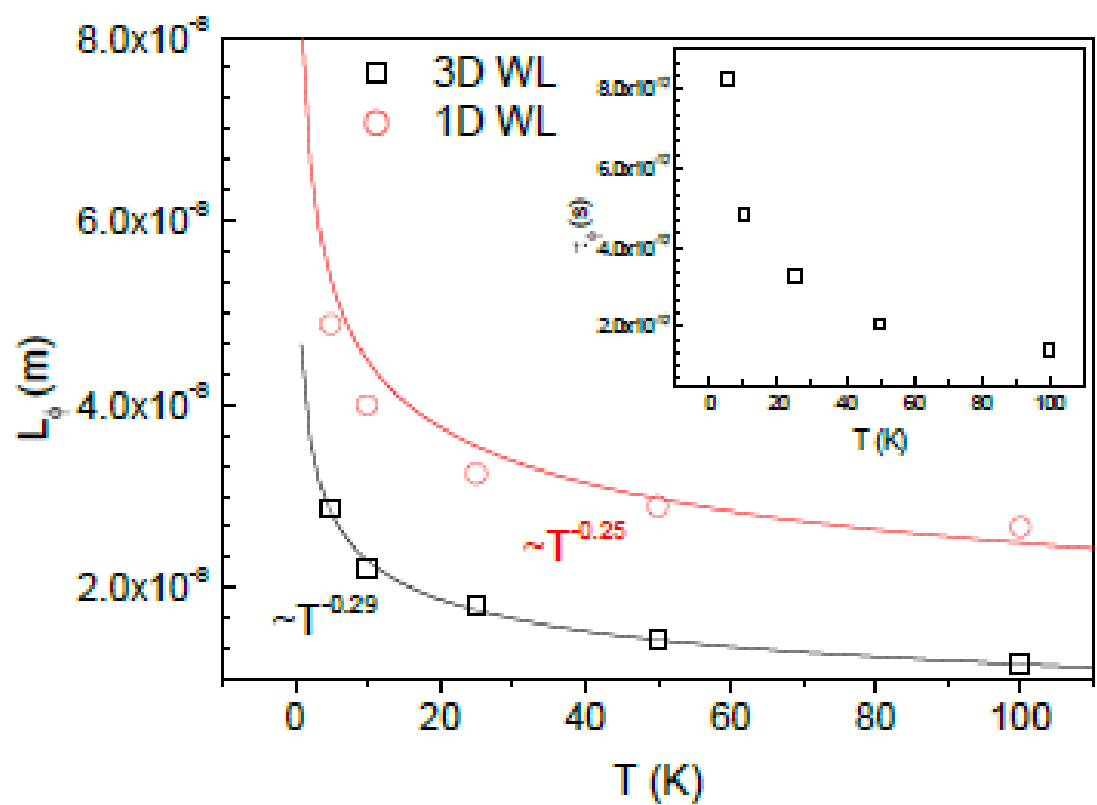

Fig. 9 / Chimowa et al.

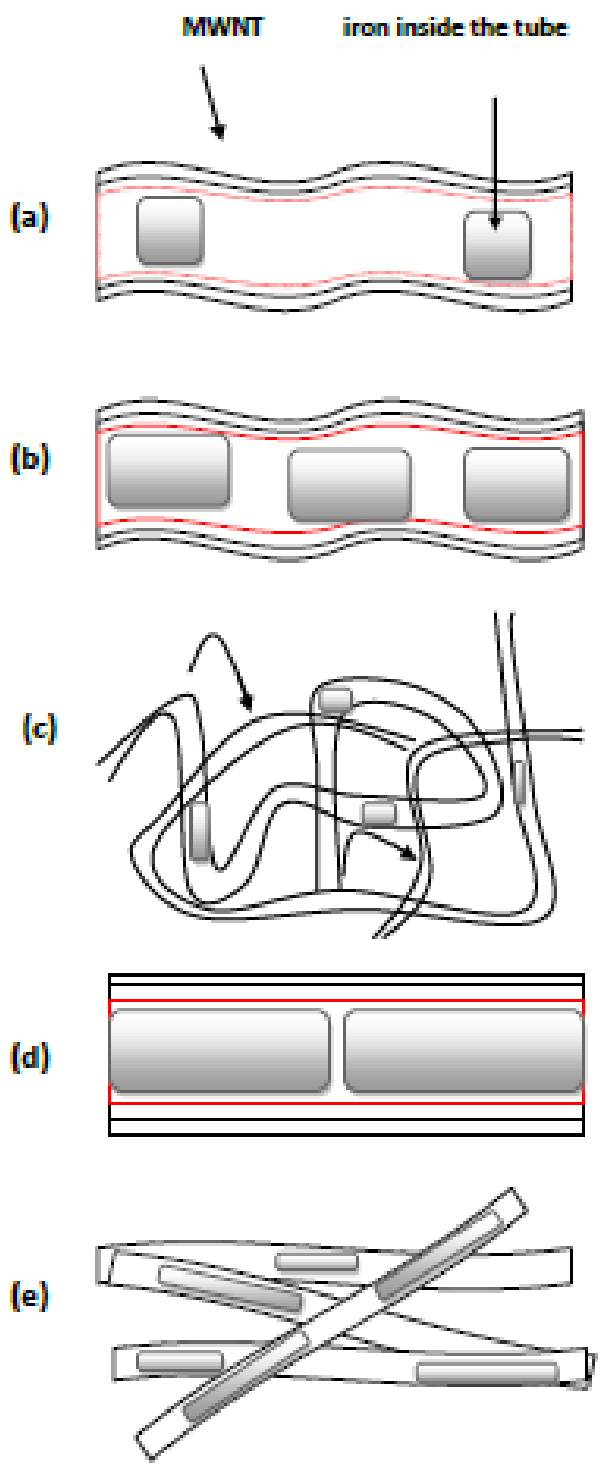

Fig. 10 (a) to (e) / Chimowa et al.